# Magnetoelastic relaxations in EuTiO$_3$

J. Schiemer[1], L. J. Spalek[2,3], S. S. Saxena[2], C. Panagopoulos[4,5], T. Katsufuji[6] and M. A. Carpenter[1]

[1] *Department of Earth Sciences, University of Cambridge - Downing Street, Cambridge CB2 3EQ, UK*
[2] *Cavendish Laboratory, University of Cambridge - Madingley Road, Cambridge CB3 0HE, UK*
[3] *Academic Centre for Materials and Nanotechnology and Faculty of Physics and Applied Computer Science, AGH University of Science and Technology - al. Mickiewicza 30, 30-059 Krakow, Poland*
[4] *Division of Physics and Applied Physics, Nanyang Technological University - 637371 Singapore, Singapore*
[5] *Department of Physics, University of Crete and FORTH - GR-71003 Heraklion, Greece*
[6] *Department of Physics, Waseda University - Tokyo 169-8555, Japan*

**Abstract** – The multiferroic properties of EuTiO$_3$ are greatly enhanced when a sample is strained, signifying that coupling between strain and structural, magnetic or ferroelectric order parameters is extremely important. Here resonant ultrasound spectroscopy has been used to investigate strain coupling effects, as well as possible additional phase transitions, through their influence on elastic and anelastic relaxations that occur as a function of temperature between 2 and 300 K and with applied magnetic field up to 14 T. Antiferromagnetic ordering is accompanied by acoustic loss and softening, and a weak magnetoelastic effect is also associated with the change in magnetization direction below ∼2.8 K. Changes in loss due to the influence of magnetic field suggest the existence of magnetic defects which couple with strain and may play a role in pinning of ferroelastic twin walls.

EuTiO$_3$ (ETO) is a perovskite structured quantum paraelectric which has the additional attribute of G-type antiferromagnetic ordering at low temperatures [1,2]. Of interest in the context of magnetoelectric materials, it shows a significant drop in the dielectric constant ($\epsilon$) at the Néel point, $T_N \approx 6$ K [3–5], due to the interaction between the spin ordering and a soft polar phonon. Normal quantum paraelectric behavior, *i.e.* smooth saturation as $T \to 0$ K, is recovered by application of magnetic fields greater than 1.5 T, with a resultant change in $\epsilon$ of up to ∼7% [4]. Shvartsman *et al.* [6] noted that the magnetoelectric magnetization comes close to that of class-leading TbPO$_4$. Strain engineering of ETO, mostly in thin films, has been theoretically predicted [7,8] and subsequently demonstrated [9] to produce a ferromagnet with a large ferroelectric polarization. This makes ETO one of the first single-phase multiferroics to display a significant polarization both magnetically and electrically.

Magnetic ordering of single-crystal samples at low temperatures involves two separate transitions [10]. The antiferromagnetic structure below $T_N$ has its easy axis of magnetization parallel to [001] of the tetragonal matrix. This gives way via a first-order transition between 2.75 and 3 K ($T_{tr}$) to a second antiferromagnetic structure with an easy plane of magnetization perpendicular to [001]. DC measurements of magnetic susceptibility under applied field have revealed an additional anomaly below $T_N$, however, which appears to involve reorientation of the direction of magnetization. The temperature for this spin-flop transition reduces from ∼5.3 K at the lowest field to 2 K at ∼0.2 T, and presumably occurs within ferroelastic twin domains which have [001] perpendicular, rather than parallel to the applied field. With increasing field the spin-flopped antiferromagnetic crystal is believed to transform to an anisotropic paramagnet, for which the critical field is ∼0.9 T at 2 K [10]. The same spin-flop and antiferromagnetic to paramagnetic transitions have also been observed in a polycrystalline sample with increasing field at 4.5 K [6]. ETO also undergoes a cubic ($Pm\bar{3}m$) to tetragonal ($I4/mcm$) ferroelastic phase transition attributed to the same octahedral tilting transition as occurs in $SrTiO_3$ (*e.g.* [10–13]) with a transition temperature, $T_c$, of 282 K suggested by a specific-heat anomaly [10,11]. Although the cubic and tetragonal structures are paramagnetic at these relatively high temperatures, it has been reported that $T_c$ changes in the presence of a magnetic field [14], implying some coupling between structural changes and spin conformation. If this is correct, it follows that the ferroelastic twin walls may also have their own distinct magnetic and dielectric properties.

Here we demonstrate the existence of coupling between strain and magnetism, obtained by correlating magnetic anomalies from SQUID data with anomalies in elastic and anelastic properties measured on the same single crystal by resonant ultrasound spectroscopy (RUS) in zero magnetic field and with applied fields of up to 14 T. Known transition behavior is examined in the context of elasticity, while evidence for additional transitions induced by increasing field is presented. We have found that applied magnetic fields and local fields from defects have effects on the properties of ETO from ∼2 K up to at least the cubic-tetragonal transition point.

The irregularly shaped single-crystal sample used in this study had approximate dimensions $2.4 \times 2.7 \times 2.3$ mm$^3$ and mass 0.0509 g. It came from one of two batches that were grown in a floating zone furnace [15] and which were also the source of crystals used by Allieta *et al.* [16], Petrović *et al.* [10] and Scagnoli *et al.* [2]. Magnetic data were collected at the Department of Engineering, University of Cambridge, on a Quantum Designs MPMS XL Squid magnetometer. RUS data were collected using a sample holder described by McKnight *et al.* [17], which was attached to the end of a stick lowered into an Oxford Instruments Teslatron PT cryostat equipped with a 14 T superconducting magnet. Fundamentals of the RUS technique are described, for example, in Migliori and Sarrao [18]. Spectra were accumulated in programmed sequences of varying temperature at constant field or varying field at constant temperature, and a delay of 20 minutes was allowed at each temperature to ensure equilibration prior to data collection. 65000 data points were collected in each spectrum, with a frequency range 300–2200 kHz. Individual resonances in the spectra were analyzed offline using the software package IGOR (Wavemetrics). The square of the frequencies, $f$, of individual resonances scale with different combinations of predominantly shear elastic constants, and the inverse mechanical quality factor, $Q^{-1} = \Delta f/f$, where $\Delta f$ is the peak width at half-height, is a measure of acoustic loss.

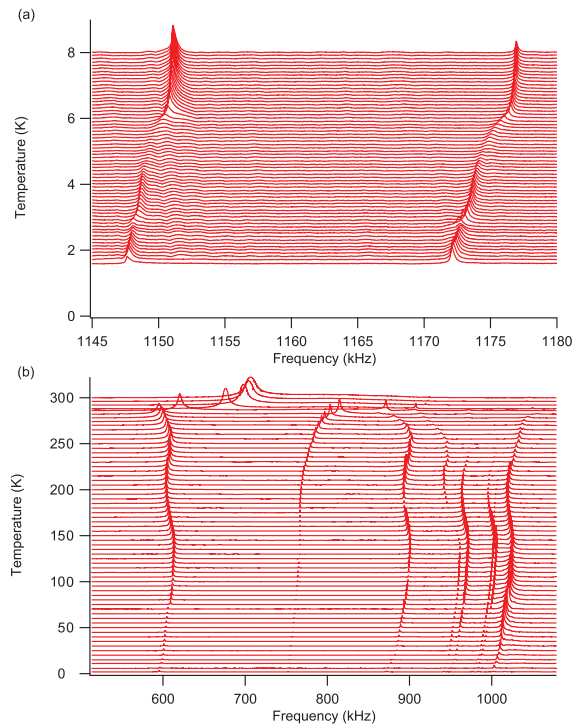

Fig. 1: (Color online) Representative stacks of RUS spectra collected during heating between (a) 1.6 K and 8 K and (b) 2 K and 300 K. The vertical axis is labeled as temperature, as the offset applied to the spectra is proportional to the temperature at which they were collected. The vertical scale for each individual spectrum is the resonance amplitude.

Representative stacks of RUS spectra are shown in fig. 1. Figure 2 contains $f^2$ and $Q^{-1}$ data from three peaks measured in 0.1 K steps during heating. They have frequencies of 994, 1172 and 1780 kHz at 1.6 K but their $f^2$ values have been scaled to 1 at this temperature to allow comparison of their evolution with increasing temperature. If there is a random orientation of domains of the tetragonal structure, the average symmetry of the crystal below $T_c$ will be cubic. In this case most of the resonances are determined predominantly by different proportions of the two shear elastic constants, "$(C_{11} - C_{12})$" and "$C_{44}$", where inverted commas are used to emphasize that these are cubic averages of the tetragonal elastic constants. Patterns of evolution through the cubic-tetragonal transition at higher temperatures [12] and through the same transition in $SrTiO_3$ suggest that the 994 kHz and 1172 kHz resonance modes can be provisionally assigned to predominately "$C_{44}$" and predominately "$(C_{11} - C_{12})$", respectively, and the 1780 kHz mode to a mixture of the two.

Two different styles of anomalies occur at $T_N$ and $T_{tr}$, for which values of 5.6 and 2.8 K taken from Petrović *et al.* [10] are shown in fig. 2. Antiferromagnetic ordering is accompanied by softening of up to ∼0.8%, although the onset of this is nearer to 6.1 K than 5.6 K, and by increased acoustic loss. There is a marked break in slope for $Q^{-1}$ at ∼6.1 K, with a peak near 5.9 K for the 994 and 1780 kHz modes and near 5.7 K for the 1172 kHz mode.

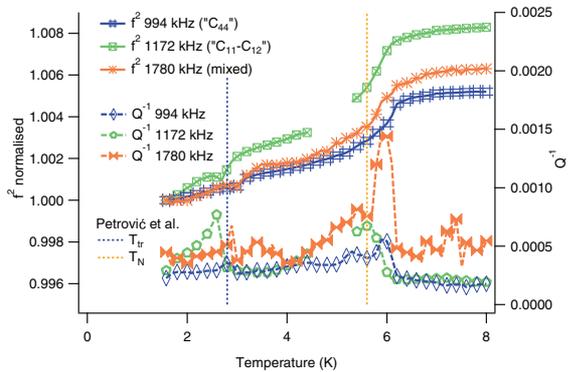

Fig. 2: (Color online) $f^2$ data for individual resonances, with absolute values given in the caption rescaled to 1 at 1.6 K, and $Q^{-1}$ data from the same resonance peaks. Vertical dotted lines mark transition temperatures taken from Petrović et al. [10] for antiferromagnetic ordering ($T_N$) and the change of magnetization direction from parallel to perpendicular to [001] ($T_{tr}$). A section with a possible artefact has been removed from the 1172 kHz peak between 4.5 K and 5.2 K. Sparse markers are used for clarity, with one marker every two data points for all traces.

Each of the three resonances has a minimum in $f^2$ and a corresponding maximum in $Q^{-1}$ near $T_{tr}$, though the exact temperature of these is at ∼2.7 K for the 1172 kHz mode and ∼2.9 K for the 994 and 1780 kHz modes.

In other systems, softening or stiffening of shear elastic constants due to antiferromagnetic ordering transitions can follow two different patterns when there is no change in crystallographic symmetry. One limiting case is seen in CoF$_2$, where some softening occurs ahead of the transition point. There is then a discontinuous softening at $T_N$ itself, which is followed by a non-linear recovery as temperature reduces further [19]. The stiffening is attributed to relaxation of the magnetic order parameter, $m$, on the same timescale as the induced strain, $e$, according to $\lambda e m^2$, where $\lambda$ represents the strength of the coupling between them. An accompanying peak in $Q^{-1}$ at $T_N$ can be related to fluctuations or to an intrinsic loss mechanism involving slowing-down of the relaxation time for $m$ as $T \to T_N$. The other limiting case is represented by YMnO$_3$ and doped BiFeO$_3$ systems, in which continuous stiffening starts at $T_N$ and increases in proportion to the square of the magnetic order parameter [20–23]. This is attributed to coupling terms of the form $\lambda e^2 m^2$ and, rather than there being a peak in $Q^{-1}$ the acoustic loss seems simply to reduce to low values below $T_N$. Coupling of this form between all strains and order parameter components is allowed, but is expected to be small and visible only when coupling with non-zero strains is extremely weak or the time scale of magnetic relaxations is greater than the time scale of the imposed strain.

Changes in the average shear elastic constants due to the magnetic ordering in ETO do not follow either of the expected patterns exactly. Softening due to biquadratic coupling is the most likely mechanism but, in the present context, the important result is that the combination of softening and acoustic loss indicates that there must be some coupling of the magnetic ordering with strain. The pattern of softening and loss near $T_{tr}$ closely resembles that due to the analogous Morin transition at ∼250 K in hematite [24]. At least some contribution to the acoustic loss may be due to movement under stress of interfaces between coexisting phases in a two-phase interval, on a time scale of ∼10$^{-6}$ s. Anomalies in the elastic properties correlate closely with anomalies in magnetic properties, as shown by measurements on the crystal held in the same orientation with respect to applied field in the SQUID magnetometer, though the onset of elastic softening occurs ∼0.4 K above $T_N$.

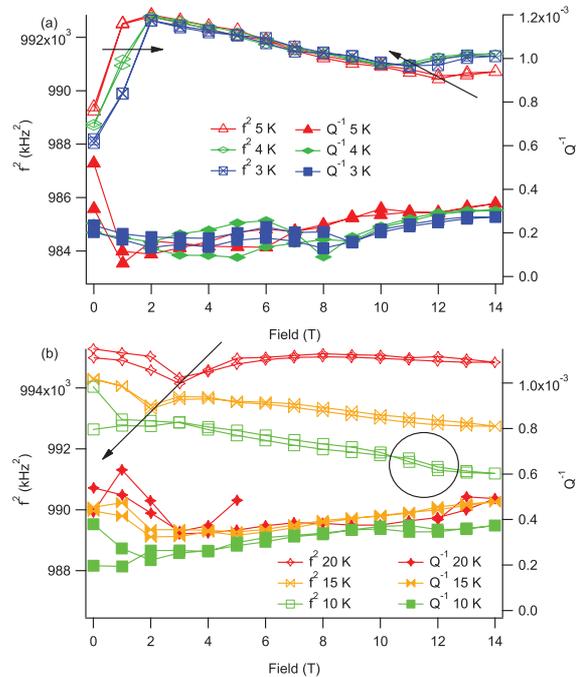

Fig. 3: (Color online) $f^2$ and $Q^{-1}$ data from the 994 kHz resonance peak as a function of magnetic field at fixed temperatures. (a) 3, 4, 5 K: anomalies occur at high and low field which vary with decreasing temperature in the direction shown by arrows. (b) 10, 15, 20 K: the circle indicates a subtle anomaly in $f^2$ at high field and 10 K, while a clear anomaly at lower field varies with decreasing temperature in the sense of the arrow. A slight hysteresis in the 10 K data shows both softening of $f^2$ and reduction in $Q^{-1}$ upon returning to zero field after application of high field.

Figure 3(a) shows variations in $f^2$ and $Q^{-1}$ for the 994 kHz resonance with changing magnetic field in 1 T steps at 3, 4 and 5 K. There are marked changes in $f^2$ for field strengths less than 2 T and, perhaps, small changes in $Q^{-1}$. These are more or less reversible at the scale of the changes in applied field and appear to correlate with the suppression of the antiferromagnetic structure under the influence of magnetic fields greater than ∼1.5 T, as previously detected by changes in heat capacity [10] and dielectric constant [4]. The stiffening with increasing field ends by 2 T, as found also for saturation of dielectric properties measured at 2 K by Katsufuji and Takagi [4]. The apparent reversibility of the changes implies that the spin-flopped and anisotropic paramagnetic structures referred to by Petrović *et al.* [10] are sustained only in an applied field. Any change in the configuration of ferroelastic twins could be reversible but it is more likely that the twin walls remain fixed while the spin configurations change within the differently oriented twin domains.

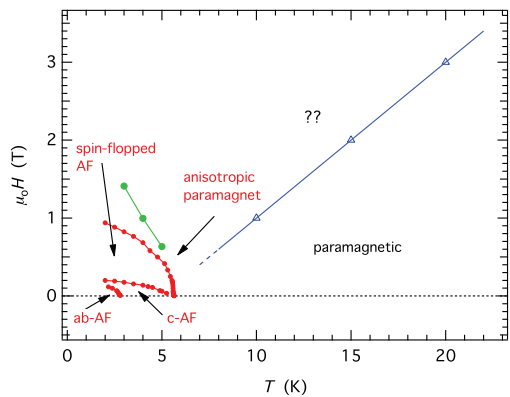

Fig. 4: (Color online) Phase diagram, as deduced from the locations of elastic anomalies detected by changing magnetic field at 2, 4, 5, 10, 15 and 20 K (blue triangles, green filled circles). Also shown are the locations of antiferromagnetic and spin reorientation transitions from ref. [10] (red filled circles joined by lines).

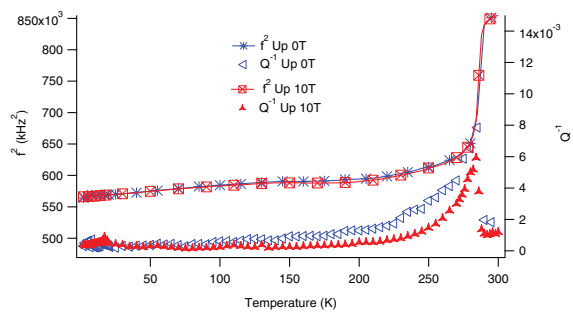

Fig. 5: (Color online) $f^2$ and $Q^{-1}$ for a resonance peak with $f = 752$ kHz at 2 K. The primary spectra were collected during heating in zero field and in an applied magnetic field of 10 T. Application of the 10 T field makes little difference to the evolution of $f^2$, even through the cubic-tetragonal transition near 285 K, but substantially reduces acoustic loss at $T \gtrsim 35$ K. Sparse markers are used for clarity, with one marker every two data points for $Q^{-1}$, one marker every three data points for $f^2$ at 0 T and one marker every four data points for $f^2$ at 10 T.

Figure 3(b) shows the influence of applied field up to 14 T at 10, 15 and 20 K. There are clear minima in $f^2$ at ∼1, 2 and 3 T, respectively, and there are also minima in $Q^{-1}$ at about the same points. These would be most simply interpreted as being indicative of additional magnetic transitions. There is a weak minimum at ∼12 T (circled) in the 10 K data, which is also seen more clearly in other peaks, and appears to match up with more overt anomalies at 10 T and above in the 3, 4 and 5 K data (fig. 3(a)). The temperature and field strength at which the breaks in slope of $f^2$ occur below 5 T are summarized in fig. 4, to which data for the boundaries between the anisotropic paramagnetic, spin-flopped antiferromagnet, *c*-AF and *ab*-AF in fig. 5a of Petrović *et al.* [10] have been added. In order to be directly comparable with the data of Petrović *et al.* [10] an approximate demagnetisation correction has been applied to the three data points at 3, 4 and 5 K (raw data 2, 1.5, 1 T; corrected data 1.41, 0.995, 0.663 T).

The locations of elastic anomalies below 2 T and 6 K in fig. 3(a) are not inconsistent with the reported changes in magnetic properties if some softening occurs ahead of the stability field of the spin-flopped antiferromagnetic structure. Petrović *et al.* [10] reported an anisotropic paramagnet as being the stable state with increasing field below 5.6 K, but it is not possible at this stage to establish what the magnetic structures might be in all of the different fields that are defined by the changes in elastic properties.

Strikingly, a straight line through the low-field data points at 10, 15 and 20 K would extrapolate to a temperature close to the Néel point in zero field. This suggests that the phase transition with increasing field could be to a structure with a different combination of magnetic order parameter components belonging to the same irreducible representation as applies to the antiferromagnetic structure below ∼6 K.

Coupling between magnetic order parameters and strain allows, in principle, the possibility of magnetic control of the ferroelastic domain structure in a manner that is analogous to the electrical-field control of ferroelastic domains described by Petrović *et al.* [10]. An electric field should induce a ferroelectric dipole with an orientation that is favored by one of the three possible ferroelastic domain orientations. If the same coupling occurs between magnetic order parameters and the tetragonal strain, an applied magnetic field should also induce poling of the ferroelastic domains. This would be detected through differences in elastic constants for poled and unpoled crystals and by a reduction in any acoustic loss related to mobility of the twin walls as their density is decreased. RUS spectra were therefore collected through the structural transition in zero field and at 10 T. Data for a resonance with $f = 751$ kHz at 1.6 K are shown in fig. 5. As described elsewhere [12], the cubic → tetragonal transition is accompanied by softening of the shear elastic constants by ∼20–30%. $f^2$ values in the stability field of the tetragonal structure are essentially the same at 0 and 10 T for this resonance and for others in the spectra, implying that there is no change in the ferroelastic domain structure. However, there is a substantial reduction of $Q^{-1}$ over almost the entire temperature range above 50 K, including above $T_c$, when the 10 T field is applied. This particular resonance mode is believed to be determined by "$C_{44}$". There is a similar reduction of $Q^{-1}$ for another mode, assigned to the influence predominantly of "$(C_{11} - C_{12})$", but the change is restricted to between ∼200 K and $T_c$. An association with "$C_{44}$" would imply coupling of strain, $e_4$, with the defect(s) responsible for the acoustic loss, which would rule out displacements of the ferroelastic twin walls as the cause. Rather, there appear to be some additional magnetic domains or defects in ETO which couple with strain at temperatures that are well above the spin ordering temperatures. These are perhaps related to the Eu$^{III}$ defects proposed by Goian et al. [13] and subsequently discussed also by Petrović et al. [10], which may also be a factor in stabilizing the incommensurate structure characterized by Kim et al. [25].

We have demonstrated magnetoelastic effects indicating that all aspects of the magnetic properties of ETO are coupled with strain, including the low-temperature antiferromagnetic phases stable below $T_N$, previously unknown magnetic structures which are stabilized in high fields and defects which persist to high temperatures. Some of the same defects may also have a role in pinning of ferroelastic twin walls. The existence of such coupling inevitably requires that strain is involved in the magnetoelectric properties, relating closely to the key results presented by Fennie et al., Ranjan et al. and Lee et al. [7–9] on the theoretical prediction and experimental realization of strongly ferroelectric, ferromagnetic epitaxial films. ETO displays aspects of three ferroic properties and clearly has a remarkable richness of phenomenological behaviour both in the bulk material and thin films.

∗ ∗ ∗

MAC acknowledges support from NERC and EPSRC (grants NE/B505738/1 and EP/I036079/1, respectively). CP acknowledges financial support in Greece through FP7-REGPOT-2012-2013-1, and in Singapore through Award No. NRF-CRP-4-2008-04 of the Competitive Research Programme. LJS acknowledges the support of the National Science Centre (NCN) through grant MAESTRO No. DEC-2012/04/A/ST3/00342. Dr Albert Migliori (Los Alamos National Laboratory) is thanked for invaluable assistance in creating the RUS system with in-situ magnetic field. Tony Dennis (University of Cambridge) collected the SQUID data.